\documentclass[aps,a4paper,10pt,twocolumn,nofootinbib]{revtex4} 
\usepackage{datetime}
\usepackage{lmodern}
\usepackage{amsmath}
\usepackage{amsfonts}
\usepackage{amsfonts}
\usepackage{mathrsfs}
\usepackage[mathscr]{euscript}
\usepackage{mathtools}
\usepackage{fancyhdr}
\usepackage{colordvi}
\usepackage{hyperref} 
\usepackage{epsfig}
\usepackage{color}
\usepackage{bm}
\usepackage[basic]{circ}

\newcommand{\XDOI}[1]{\href{http://dx.doi.org/#1}{doi:#1}}
\newcommand{\XARXIV}[1]{\href{http://arxiv.org/abs/#1}{arXiv:#1}}
\newcommand{\XWEB}[1]{\href{#1}{#1}}
\renewcommand*\rmdefault{ptm} 

\def\overstrike#1#2{{\setbox0\hbox{$#2$}\hbox to \wd0{\hss
    $#1$\hss}\kern-\wd0\box0}}

\renewcommand{\Vec}[1]{\boldsymbol{#1}}  

\newcommand{\pTime}{t}

\newcommand{\pSpaceZ}{z}

\newcommand{\pPderivT}{\partial_{\pTime}}   
\newcommand{\pPderivZ}{\partial_{\pSpaceZ}} 

\newcommand{\pContour}{C}
\newcommand{\pXdline}{\Vec{dl}}

\newcommand{\pArea}{S}
\newcommand{\pSurface}{S}
\newcommand{\pXdsurface}{\Vec{dS}}
\newcommand{\pLightspeed}{c}

\newcommand{\pPermittivity}{\epsilon}   

\newcommand{\pPermeability}{\mu}         

\newcommand{\pXemXelectric}{E}                      
\newcommand{\pXemXelectricv}{\Vec{\pXemXelectric}}  
\newcommand{\pXemXdisplacement}{D}                         
\newcommand{\pXemXdisplacementv}{\Vec{\pXemXdisplacement}} 

\newcommand{\pXemXmagnetic}{B}                           
\newcommand{\pXemXmagneticv}{\Vec{\pXemXmagnetic}}       
\newcommand{\pXemXmagstrength}{H}                        
\newcommand{\pXemXmagstrengthv}{\Vec{\pXemXmagstrength}} 

\newcommand{\pCharge}{q}                   
\newcommand{\pCurrent}{J}                  
\newcommand{\pCurrentv}{\Vec{\pCurrent}}   
\newcommand{\pXemf}{\mathscr{V}}           
\newcommand{\pXmagflux}{\Phi}              

\newcommand{\pBfield}{\pXemXmagnetic}

\newcommand{\pEfieldv}{\pXemXelectricv}
\newcommand{\pDfieldv}{\pXemXdisplacementv}

\newcommand{\pBfieldv}{\pXemXmagneticv}
\newcommand{\pHfieldv}{\pXemXmagstrengthv}

\newcommand{\pWork}{W}   

\newcommand{\pForce}{F}              
\newcommand{\pForcev}{\Vec{\pForce}} 

\newcommand{\pMomentum}{p}                  
\newcommand{\pMomentumv}{\Vec{\pMomentum}}  

\newcommand{\pSpeed}{v}                  
\newcommand{\pVelocity}{\Vec{\pSpeed}}   

        \DeclareMathOperator{\grad}{\nabla}
        \DeclareMathOperator{\cross}{\times}

\def\EMF{\pXemf}

\def\Pcur{P_{\textrm{cur}}} 
\def\Pbar{P_{\textrm{bar}}} 
\def\sbar{v}                
\def\vbar{\Vec{\sbar}}      
\def\scharge{u}             
\def\vcharge{\Vec{\scharge}}
\def\QMdensity{\rho}        
\def\Qdensity{\sigma}       
\def\Fbarv{\Vec{F}_{\textrm{bar}}} 
\def\Fchargev{\Vec{F}_q}    
\def\sCurrent{J}            
\def\vCurrent{\Vec{\sCurrent}}        
\def\Mbar{M}                
\def\Qtotal{Q}              
\def\Qbar{q}                

\def\Ltotal{L_\textrm{tot}} 
\def\Lbar{L_\textrm{bar}}   
\def\Larm{L}                

\def\Ushape{\textsf{U}}

\newdateformat{yymmdddate}{\THEYEAR/\twodigit{\THEMONTH}/\twodigit{\THEDAY}}

\definecolor{XRED}{rgb}{0.71, 0.01, 0.01}
\definecolor{XBLUE}{rgb}{0.01, 0.01, 0.71}
\def\UPDATED#1{#1}
\def\UPDATES#1{#1}

\definecolor{XGREEN}{rgb}{0.01, 0.31, 0.01}

\begin{document}
\title{Faraday's Law and Magnetic Induction: 
cause and effect}
\author{Paul Kinsler$^{1,2}$}
\homepage[]{https://orcid.org/0000-0001-5744-8146}
\email[\hphantom{.}~]{Dr.Paul.Kinsler@physics.org}
\affiliation{$^1$
  Physics Department,
  Lancaster University,
  Lancaster LA1 4YB,
  United Kingdom.}
\affiliation{$^2$
  Cockcroft Institute, Keckwick Lane,
  Daresbury,
  WA4 4AD,
  United Kingdom.}

\lhead{\includegraphics[height=5mm,angle=0]{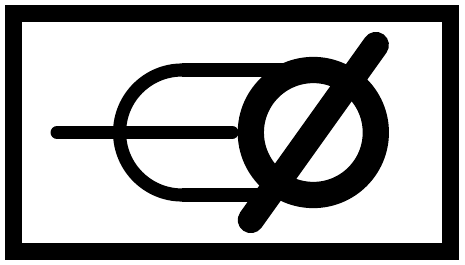}~~FARADIN}
\chead{Faraday's Law of Induction}
\rhead{
\href{mailto:Dr.Paul.Kinsler@physics.org}{Dr.Paul.Kinsler@physics.org}\\
\href{http://www.kinsler.org/physics/}{http://www.kinsler.org/physics/}
}

\begin{abstract}

Faraday's Law of induction is often stated  as 
 ``a change in magnetic flux causes an electro-motive force (EMF)'';
 or, 
 more cautiously,
 ``a change in magnetic flux is associated with an EMF''.
It is as well that the more cautious form exists, 
 because the first ``causes'' form 
 can be shown to be
 incompatible with the usual expression
 $\pXemf = -  \pPderivT  \pXmagflux$,
 where $\pXemf$ is EMF,
 $\pPderivT$ a time derivative,
 and $\pXmagflux$ the magnetic flux.
This is not, 
 however, 
 to deny the causality as reasonably inferred from experimental observation --
 it is the \emph{equation} for Faraday's Law of induction
 which does not represent the claimed cause-and-effect relationship.
Unusually, 
 in this induction scenario, 
 the apparent experimental causality 
 does not match up with that of the mathematical model. 
Here we investigate a selection of different approaches, 
 trying to see how an explicitly causal mathematical equation,
 which attempts to encapsulate the 
 ``a change in magnetic flux causes ...'' idea, 
 might arise.
\UPDATES{We see that although it is easy to find
 mathematical models 
 where changes in magnetic flux or field 
 have an effect on the electric current, 
 the same is not true for the EMF.}

\end{abstract}


\date{\today}
\maketitle
\thispagestyle{fancy}

%

%
\section{Introduction}\label{S-intro}

Michael Faraday was the first person to publish experimental results 
 on electromagnetic induction, 
 a phenomenon which is frequently described 
 as a current being induced when the magnetic flux through a conducting coil
 is changed.

The phenomenon is compactly described by a mathematical model
 relating the ``electro-motive force'' (EMF)
 $\pXemf$ around a closed circuit
 to the temporal changes in magnetic flux $\pXmagflux$ -- 
 the integral of the magnetic field over the surface delineated by that
 closure.
This Faraday's law of induction
 can be directly related to the Maxwell-Faraday vector equation 
 \cite[p.160]{Bleaney2-EM}.
Using the notation where $\pPderivT \equiv d/dt$, 
 the equations are written as
~
\begin{align}
  \pXemf 
&=
 -
  \pPderivT  \pXmagflux
\quad \leftrightarrow \quad
  \grad \cross \pEfieldv
=
 -
   \pPderivT \pBfieldv 
,
\label{eqn-inductorizing}
\end{align}
 where $\pEfieldv$ is electric field and $\pBfieldv$ is magnetic field.

However, 
 a fundamental law of causality demands that 
 effects cannot occur before their causes; 
 and this expectation has mathematical consequences.
These consequences give rise in particular
 to the famous Kramers Kronig relations 
 \cite{Bohren-2010ejp,LandauLifshitz,KKinOMR},
 which can be used to constrain both the spectral properties
 of physical quantities and the models used to describe them.

Here we take a local view of causality, 
 rather than a global one, 
 and use the time domain most natural to expressions of causal behaviour.
In this case a scheme exists \cite{Kinsler-2011ejp}
 for attributing specific cause and effect roles
 to the components of mathematical models that are
 based on temporal differential equations\footnote{See also e.g. 
  \cite{Corduneanu-FECAUSAL,Corduneanu-2000jde} for mathematical details.}.
Most simply, 
 this means that 
 in any equation of the mathematical model 
 it is the highest-order time derivative term
 which is to be regarded as the \emph{effect}, 
 with the other terms as causes; 
 but note that it can also be applied 
 to systems of coupled equations
 (e.g.  Maxwell's equations \cite[Sec. V]{Kinsler-2011ejp}, 
 as well as later in this paper in Subsec. \ref{S-forcelaw-example}).
As a simple example, 
 since Newton's second law can be written as
 $\pPderivT \pVelocity = \pForcev/m$,
 then according to \cite{Kinsler-2011ejp}, 
 we can state that
 the force $\pForcev$ causes a change in the velocity $\pVelocity$ of a mass $m$.
The use of this interpretation
 is particularly natural when considering 
 the computational solution of dynamic physical models
 (see \cite[Appendix A]{Kinsler-2011arxiv-how2c}, 
 and where the main focus of attention
 is what happens
 as the current state of the system is updated
 \cite{Kinsler-2014arXiv-negfreq,Kinsler-2018jo-d2owe}
 (i.e., integrated forwards in time), 
 or as the universe extends itself into its ``future''
 (as in e.g. the causal set approach \cite{Rideout-S-2000prd}).

Note that other discussions of and approaches to
 causality in electrodynamics relevant
 to Faraday's Law do exist
 \cite{Jefimenko-2004ejp,Hill-2010pt,Savage-2011arXiv-cem}.
In particular, 
 one significant school of thought prefers to relate everything
 back to the original source terms (see e.g. \cite{Jefimenko-2004ejp}, 
 and remarks in \cite[Appendix B]{Kinsler-2011arxiv-how2c}).
Unfortunately, 
 since Faraday's Law contains no explicit references to source charges, 
 we cannot analyse it in regards to cause and effect 
 on that basis -- 
 we would need a different model that did refer to the source charges 
 and their motions
 (e.g. see \cite{Boyer-2015ajp}).

\emph{Consequently, 
 in what follows we choose to only use
 the local and
 Kramers Kronig compatible definition \cite{Kinsler-2011ejp}
 for what constitutes a causal interpretation of a mathematical model.}

Using the local view of causality
 described in \cite{Kinsler-2011ejp},
 since \eqref{eqn-inductorizing}
 has no time derivatives applied to $\pXemf$,
 but one time derivative applied to $\pXmagflux$, 
 this means that 
 the EMF $\pXemf$ must be considered a \emph{cause},
 where the change in flux $\pPderivT  \pXmagflux$
 is its  \emph{effect}.
As a consequence,
 it would be more intuitive to rewrite \eqref{eqn-inductorizing}
 instead as 
~
\begin{align}
  \pPderivT  \pXmagflux
&=
 -
  \pXemf 
\quad \leftrightarrow \quad
   \pPderivT \pBfieldv 
=
 -
  \grad \cross \pEfieldv
,
\label{eqn-induction}
\end{align}
 where we are likewise compelled to describe the 
 spatial gradients of the electric field $\grad \cross \pEfieldv$
 as a cause,
 and the resulting temporal changes in field $\pBfieldv$ as its effect.
Somewhat disturbingly, 
 this now means we are unable to interpret our \emph{mathematical model}
 of Faraday's law of induction as our preferred causal statement:
 i.e.
 where changes in flux induce an EMF (and hence drive currents)
 \cite{Thide-EMFT,Jackson-ClassicalED,RMC}.
However,
 based on our mathematical model,
 we are still able to make the weaker statement
 where 
 the two are merely equated or associated with one another\footnote{When 
  looking up the definition of ``induce'' we see that it is 
  in essence the same as ``causes''.
 Some 
  instances of the various phrasings of
  ``cause'', ``induce'',  and ``associated with''
  from the literature are discussed in the appendix.}.

~

{\color{XGREEN}{\setlength{\fboxrule}{1.75pt}
\fbox{\parbox{0.90\columnwidth}{
\textbf{How to be causal:}
Consider the simple model equation $\pPderivT R(t) = Q(t)$, 
 which under the interpretation used here
 is interpreted as ``$Q$ causes changes in $R$''.
Using the mathematical definition of the derivative, 
 at some arbitrary time $t_0$
 the model equation is 
~
\begin{align}
  \lim_{\delta \rightarrow 0}
  \frac{R(t_0+\delta) - R(t_0-\delta)}
       {2\delta}
&=
 Q(t_0)
.
\end{align}
Just before we take the limit, 
 and for some small enough value of $\delta$,
 we have 
~
\begin{align}
  \frac{R(t_0+\delta) - R(t_0-\delta)}
       {2\delta}
&=
 Q(t_0)
.
\label{eqn-box-nolimit}
\end{align}
Now let us try to use this as a predictor of the future.
We immediately see that the latest-time 
 (or most future-like)
 quantity is $R(t_0+\delta)$, 
 so 
 we rearrange to put this on the left, 
 and get
~
\begin{align}
  R(t_0+\delta)
&=
  R(t_0-\delta)
 +
  2\delta Q(t_0)
.
\end{align}
Here, 
 we see that to update our knowledge of $R$
 to its next value at $t+\delta$
 requires (a) past knowledge of $R$ (i.e. $R(t_0-\delta)$)
 and (b) current knowledge of $Q$ (i.e. $Q(t_0$)).
Thus it is clear that the value of $Q$ causes changes in $R$.

If we instead wanted to calculate $Q$, 
 hoping to say something like ``$R$ causes $Q$'',
 we would first rearrange \eqref{eqn-box-nolimit} into
~
\begin{align}
 Q(t_0)
&=
  \frac{R(t_0+\delta) - R(t_0-\delta)}
       {2\delta}
,
\end{align}
 but will immediately see that $Q(t_0)$
 depends on a future value of $R$, 
 namely $R(t_0+\delta)$ -- 
 and a dependence on future values is incompatible 
 with standard notions of causality.
}}
}}

~

This situation suggests that 
 \eqref{eqn-induction}, 
 despite its utility, 
 is simply not a good causal representation of the experiment
 we had in mind.
To resolve this 
 we need to to make a clear distinction between 
 the causality apparent in an experiment, 
 and the causal interpretation of the mathematical model that is --
 necessarily --
 only an approximation to that experiment.
While we would usually hope that these agree, 
 it seems that in the case of Faraday's law they do not, 
 \UPDATED{and in Sec. \ref{S-experiment}
 we discuss why (or how)
 this can be.}


Clearly, 
 it would be desirable for
 our mathematical description of induction to explicitly show
 how EMF or electrical current could be generated in a conductor, 
 whether by the varying properties of the magnetic field, 
 or by motion of the conductor within those fields.
In what follows we try three (non-relativistic) approaches
 to finding a mathematical model which describes
 how some property or behaviour of the magnetic flux
 (or perhaps just the magnetic field)
 is explicitly attributable as being the \emph{cause} 
 of an EMF.

The first, 
 in section \ref{S-faraday}
 is based on the Maxwell-Faraday equation, 
 and fails.
The second, 
 in section \ref{S-ampere}, 
 is based instead on the Maxwell-Ampere equation,
 and achieves that essential goal, 
 but perhaps not in a very satisfactory way.
The third,
 in section \ref{S-forcelaw},
 is based on the Lorentz force law, 
 and we try both a simple abstract calculation 
 as well as a specific setup, 
 common in many undergraduate courses, 
 and derive a model for the current due to magnetic fields and motion.
Finally, 
 in section \ref{S-conclusion},
 we conclude.

This undergraduate level presentation 
 discussing magnetic induction
 highlights the distinction between any inferred experimental causality
 and that encoded in its mathematical model.
It shows that the apparently useful definition of an EMF 
 actually works \emph{against} our attempts to generate a model
 where it is induced by changes in magnetic flux.
In contrast, 
 an alternative focus on the electrical currents generated by motion 
 and or magnetic field variation 
 has no such limitations.

%
\section{Experiment}\label{S-experiment}

\begin{figure}
  \begin{center}
    \resizebox{0.95\columnwidth}{!}{
    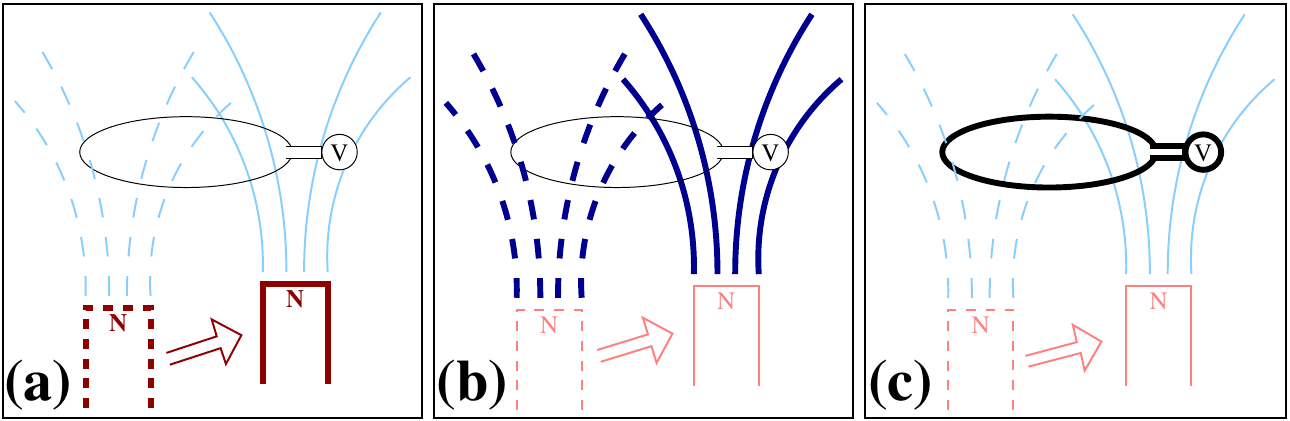}
  \end{center}
\caption{
An experiment:
 (a) a magnet (its north pole marked ``N'')
 with velocity $\Vec{v}(t)$ passes near a wire loop
 with a voltmeter {\textcircled{v}},
 so that 
 (b) its magnetic field $\pBfieldv(t,\Vec{r})$ moves also, 
 leading to 
 (c) an EMF $\EMF$ appearing around the loop.
Although going from (a) to (b) is simple to model, 
 being just a translation of the magnetic field of the magnet, 
 there is much microscopic detail in (c), 
 where those fields
 interact with the charges in the loop, 
 inducing and altering currents, 
 until the eventual effect on the EMF.
}
\label{fig-experiment}
\end{figure}

\UPDATED{Consider an example experiment
 consisting of two pieces of apparatus, 
 (i) a magnet, and (ii) a loop or coil with a voltmeter attached, 
 as depicted in fig. \ref{fig-experiment}.
When we move the magnet in the vicinity of the loop, 
 we see a voltage induced.
On the basis of this experimental experience, 
 it is therefore quite natural --
 and indeed accurate -- 
 to infer that moving the magnet caused a voltage (or EMF) appear.
If also a student of physics, 
 we might then also relate the motion of the magnet
 to the amount of magnetic flux crossing the loop, 
 and so say our experiment indicates that ``a change in flux 
 causes an EMF''.}

\UPDATED{The ``experimental causality'' just described 
 is not controversial.
However, 
 what \emph{is} under examination here is the 
 conflation of this experimental experience of causality
 with any causal interpretation of 
 the mathematical expression of Faraday's Law, 
 i.e. \eqref{eqn-induction}.}

\UPDATED{For a better understanding 
 we need to start by being clearer about \emph{why} we might deduce that the 
 voltage was caused by our actions in moving the magnet.
Notably, 
 we should realise that this is not merely because we saw
 a finite voltage registered on the voltmeter, 
 since it might simply have been poorly zeroed
 and always have some finite reading,
 or offset by the effects of some external field
 irrelevant to the induction experiment.
Instead, 
 we say ``caused''
 because after (and as) we moved the magnet near the loop, 
 the voltage \emph{changed}.
The \emph{change} in voltage
 is the visible effect that leads us to look for its cause.
At the very least, 
 then, 
 we should want our causal model of induction to 
 describe how changes in voltage 
 (``EMF'') occur due to external stimuli.}

\UPDATED{Let us now try to mathematically codify our experiment, 
 allowing for the pre-defined input
 (the motion of the magnet), 
 while looking for the resulting effect on the output (the EMF $\EMF$).
Assuming for simplicity a fixed-strength magnet, 
 with an equally fixed orientation,
 the sole input to our model only needs to be its velocity $\Vec{v}(t)$.
This $\Vec{v}(t)$ is the cause; 
 it is something we specify in the experimental design 
 and then implement.
We are hoping --
 or expecting --
 this cause to give rise to a change.
This change should be visible as an effect on the measured EMF, 
 so that at one moment it might be zero, 
 but the next it will be different -- 
 i.e. it should change with time, 
 and somehow as a result of the cause $\Vec{v}(t)$.
Thus we might write 
~
\begin{align}
  \pPderivT \EMF 
&=
  \mathscr{L}(\Vec{v}(t))
,
\label{eqn-inductoform}
\end{align}
 where $\mathscr{L}()$ is a function or operator 
 understood to contain
 all the necessary information about electromagnetism, 
 the field pattern produced by the magnet,
 the Lorentz force law on charges, 
 and circuit theory.
}

\UPDATED{We can at a glance now see that the 
 structure of the mathematical model 
 \eqref{eqn-inductoform}
 for our experiment is not of the same form
 as Faraday's Law: 
 \eqref{eqn-induction} only explicitly refers
 to changes in flux $\pPderivT \pXmagflux$, 
 and not to changes in EMF.
Indeed, 
 this should not need be surprising, 
 since \eqref{eqn-induction}
 is not even an attempt at a microscopic model of the experimental process, 
 where the moving magnet 
 produces a changing electromagnetic field distribution,
 which then goes on to 
 influence the motion of electrons, 
 producing or changing electrical currents
 (e.g. \cite{Boyer-2015ajp}), 
 and thus subsequently giving rise to an EMF.
Since so much experimental detail is omitted from \eqref{eqn-induction}, 
 it is not at first sight clear
 why we should expect the two to match at all, 
 let alone perfectly.}

\UPDATED{Nevertheless, 
 in what follows we will attempt to find a Faraday-like law 
 that matches better to our experimental concept.}


%
\section{Maxwell-Faraday Equation}\label{S-faraday}

First let us consider the 
 standard derivation of Faraday's law of induction
 starting with the curl Maxwell's equation;
 note however some approaches take the reverse approach,
 and 
 start with induction and reduce it to the curl equation.

\begin{figure}
  \begin{center}
    \resizebox{0.95\columnwidth}{!}{
    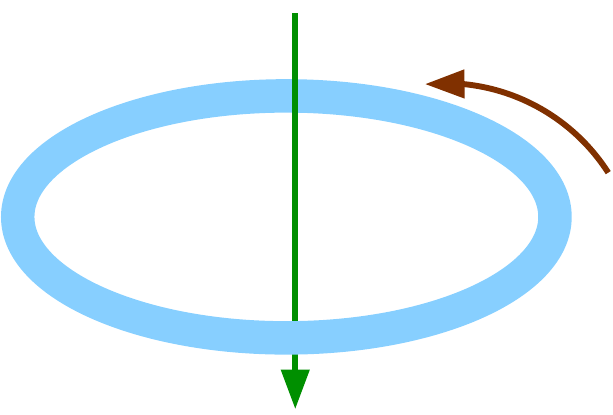}
  \end{center}
\caption{
Diagram showing the configuration and fields
 used for the induction equation 
 based -- 
 as usual -- 
 on the Maxwell-Faraday equation in fields $\pEfieldv$ and $\pBfieldv$.
The loop that allows us to decide on the integration contour $\pContour$
 is shown in blue, 
 and the surface of integration $\pSurface$ is its cross-sectional disc.
}
\label{fig-faraday-OF}
\end{figure}

\begin{figure}
  \begin{center}
    \resizebox{0.75\columnwidth}{!}{
    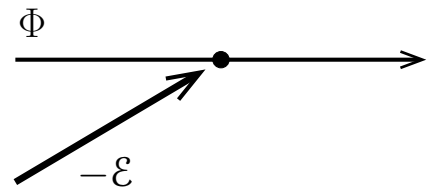}
  \end{center}
\caption{
Causal diagram \cite{Kinsler-2015arXiv-dicaus}
 for the standard Maxwell-Faraday based
 induction equation, 
 showing how the EMF ${\pXemf}$ 
 causes changes in the magnetic flux $\pXmagflux$,
 as per \eqref{eqn-inductorizing} or \eqref{eqn-MF-induction}.
}
\label{fig-faraday-MFdicaus}
\end{figure}

We start with a loop of material,
 and regard that loop as the boundary $\pContour$ 
 of some enclosed surface $\pSurface$, 
 and treat the electric and magnetic fields 
 $\pEfieldv$ and $\pBfieldv$,
 as in fig. \ref{fig-faraday-OF}
The derivation proceeds by starting with the 
 Maxwell-Faraday equation, 
 taking a surface integral of both sides, 
 and then 
 applying Stokes' theorem to the LHS
 in order to convert that surface integral
 into a line integral over elements $\pXdline$ along $\pContour$:
~
\begin{align}
  \pPderivT \pBfieldv 
&=
 -
  \grad \cross \pEfieldv
\\
  \int_\pSurface \pXdsurface \cdot
  \pPderivT \pBfieldv 
&=
 -
  \int_\pSurface \pXdsurface \cdot
  \grad \cross \pEfieldv
\\
  \pPderivT 
  \int_\pSurface \pXdsurface \cdot
  \pBfieldv 
&=
 -
  \oint_\pContour \pXdline \cdot
  \pEfieldv
\label{eqn-MF-inductionEMF}
\\
  \pPderivT 
  \pXmagflux
&=
 -
  {\pXemf}
.
\label{eqn-MF-induction}
\end{align}
The causal interpretation of this equation
 must be \cite{Kinsler-2011ejp} 
 that the EMF ``causes'' changes in flux $\pXmagflux$ --
 we cannot in fact take 
 the common interpretation that changes in flux induce (or cause) an EMF.
Note that this is not an interpretation
 of experimental facts or observations, 
 but one based on the mathematical model.
A causal diagram \cite{Kinsler-2015arXiv-dicaus}
 for the mathematical model is shown in fig. \ref{fig-faraday-MFdicaus}.

Further, 
 note that the EMF, 
 i.e. 
 $\pXemf = \oint_\pContour \pXdline \cdot \pEfieldv$,
 which is supposed to represent the current-generating potential
 induced in the wire,
 is not calculated as we perhaps might have expected:
 i.e.
 by following a single charge on its path around a conducting loop 
 to work out the voltage difference traversed.
Instead what we have done is
 consider the forces applied
 to a continuous line of infinitesimal charges on the loop
 \emph{at a single instant},
 and integrated those forces.
Without a model for the motion of these charges, 
 what we have modelled is more like a dielectric ring, 
 with \emph{bound} charges and currents tied to their own locations, 
 and not a conducting wire loop with freely moving charges and currents
 \cite{Kumar-2017ejp}.

Of course, 
 from a practical perspective Faraday's Law of induction
 derived above and shown in eqn. \eqref{eqn-inductorizing}
 is an exceedingly useful expression,
 and it is compatible with experimental measurements.
Notably, 
 in an electric motor or generator, 
 the speed of the axle's rotation
 supplies us a natural angular frequency $\omega$,
 so that the time derivative of flux becomes a simple 
 multiplication by $\omega$; 
 and further,
 in such a quasi-stationary case, 
 attribution of causality becomes unimportant.
However, 
 its utility is not any guarantee that it will supply
 a mathematical model encapsulating the casual properties 
 \emph{that we might hope for}
 for the generation of \UPDATES{EMF. 
Neverthless, 
 we will see later in section \ref{S-forcelaw}
 that this difficulty is not present
 when finding models that show effects on the current
 as a result of changes in magnetic field or flux.}

%
\section{Maxwell-Ampere Equation}\label{S-ampere}

In section \ref{S-faraday} we saw that the Maxwell-Faraday equation
 did not provide the causal interpretation we wanted.
Consequently,
 let us try an alternative derivation for a model of induction
 starting with the  curl Maxwell's equation
 that represents Ampere's Law.
The aim here is to show how some suitable property of the magnetic field --
 hopefully related to a magnetic flux --
 explicitly causes an EMF to change.
This would mean we can address the generation of an EMF
 from nothing, 
 and use that to motivate how an induced current will appear.
Since we have seen in the preceeding section
 that according to \cite{Kinsler-2011ejp},
 the
 Maxwell-Faraday equation is the wrong way around 
 to motivate a ``magnetic fields causing (inducing) current'' picture,
 the Maxwell-Ampere equation with its reversed roles
 for electric and magnetic contributions 
 may seem more promising.

%

\begin{figure}
  \begin{center}
    \resizebox{0.95\columnwidth}{!}{
    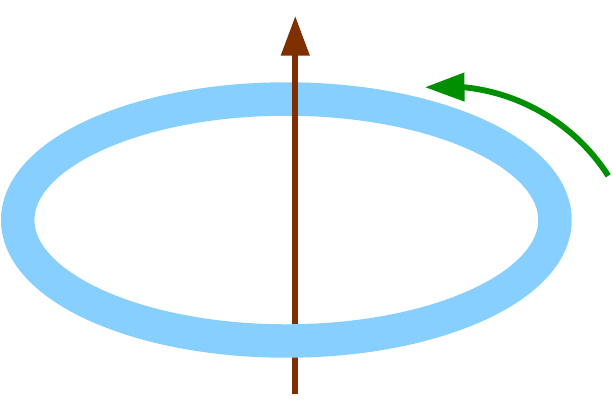}
  \end{center}
\caption{
Diagram showing the configuration and fields
 used for the induction equation 
 based 
 on the Maxwell-Ampere equation in fields $\pDfieldv$ and $\pHfieldv$.
The loop that allows us to decide on the integration contour $\pContour$
 is shown in blue, 
 and the surface of integration $\pSurface$ is its cross-sectional disc.
}
\label{fig-faraday-OA}
\end{figure}

Again we take a loop containing charges,
 with enclosed surface $\pSurface$
 and boundary $\pContour$, 
 and treat the electric displacement and magnetic induction fields 
 $\pDfieldv$ and $\pHfieldv$,
 as in fig. \ref{fig-faraday-OA}.
The derivation proceeds by starting with the 
 Maxwell-Ampere equation, 
 taking a surface integral of both sides, 
 and then 
 applying Stokes' theorem to the RHS
 in order to convert that surface integral
 into a line integral:
~
\begin{align}
  \pPderivT \pDfieldv
&=
  \grad \cross \pHfieldv
 -
 \pCurrentv
\\
  \int_\pSurface \pXdsurface \cdot
  \pPderivT \pDfieldv
&=
  \int_\pSurface \pXdsurface \cdot
  \left[
    \grad \cross \pHfieldv
  \right]
 -
  \int_\pSurface \pXdsurface \cdot
  \pCurrentv
\\
  \pPermittivity
  \pPderivT 
  \int_\pSurface \pXdsurface \cdot
  \pEfieldv
&=
  \pPermeability^{-1}
  \oint_\pContour \pXdline \cdot
  \pBfieldv
 -
  \int_\pSurface \pXdsurface \cdot
  \pCurrentv
,
% ...
\end{align}
 where $\pCurrentv$ is the electric current.

From this we can 
 straightforwardly see that 
 neither an EMF or a flux $\pXmagflux$ emerges.
But some further thought leads us to 
 more success by means of taking the curl of both sides
 \emph{before} applying the surface integral.
With $\pPermittivity$ being the permittivity,
 and $\pPermeability$ the permeability, we have:
~
\begin{align}
    \grad
   \cross 
    \pPderivT 
    \pDfieldv
&=
  \grad \cross
  \grad \cross
  \pHfieldv
 -
  \grad \cross
  \pCurrentv
\\
  \pPermittivity
  \int_\pSurface \pXdsurface \cdot
  \left[
    \grad \cross
    \pPderivT 
    \pEfieldv
  \right]
&=
  \pPermeability^{-1}
  \int_\pSurface \pXdsurface \cdot
  \left[
    \grad \cross
    \grad \cross
    \pBfield
  \right]
\nonumber
\\
&\qquad
 -
  \int_\pSurface \pXdsurface \cdot
  \left[
    \grad \cross
    \pCurrentv
  \right]
\\
  \pPermittivity
    \pPderivT 
  \oint_\pContour \pXdline \cdot
    \pEfieldv
&=
  \pPermeability^{-1}
  \int_\pSurface \pXdsurface \cdot
  \left[
    \grad \cross
    \grad \cross
    \pBfieldv
  \right]
\nonumber
\\
&\qquad
 -
  \oint_\pContour \pXdline \cdot
    \pCurrentv
\\
    \pPderivT 
  \oint_\pContour \pXdline \cdot
    \pEfieldv
&=
  \left(
  \pPermeability   \pPermittivity
  \right)^{-1}
  \int_\pSurface \pXdsurface \cdot
  \left[
    \grad \cross
    \grad \cross
    \pBfieldv
  \right]
\nonumber
\\
&\qquad
 -
  \pPermittivity^{-1}
  \oint_\pContour \pXdline \cdot
    \pCurrentv
,
\end{align}

This equation does now give us
 an EMF from the LHS line integral, 
 calculated just as in section \ref{S-faraday}.
Since $\grad \cdot \pBfieldv = {0}$, 
 we can therefore write
~
\begin{align}
    \pPderivT 
  \pXemf
&=
  \pLightspeed^2
  \int_\pSurface \pXdsurface \cdot
  \left[
    \grad^2
    \pBfieldv
  \right]
\nonumber
\\
&\qquad
 -
  \pPermittivity^{-1}
  \oint_\pContour \pXdline \cdot
    \pCurrentv
.
\label{eqn-FaradayAmpere}
\end{align}
This result is an induction-like equation
 which we can give
 a well defined causal interpretation where the EMF is caused:
 being that 
 \emph{spatial} variations in $\pBfieldv$ 
 (or, when integrated, spatial variations in flux),
 drive changes in EMF; 
 as do currents.

\begin{figure}
  \begin{center}
    \resizebox{0.75\columnwidth}{!}{
    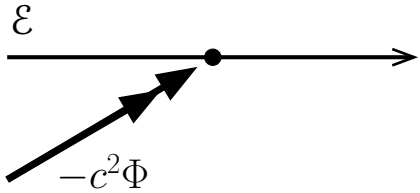}
  \end{center}
\caption{
Causal diagram \cite{Kinsler-2015arXiv-dicaus}
 for the Maxwell-Ampere based
 induction equation, 
 i.e. \eqref{eqn-MAdicaus},
  showing how flux gradients cause changes in EMF.
The double \emph{filled} arrows here indicate the double spatial derivative
 applied to $\pBfield_z$ and hence the flux $\pXmagflux$.
}
\label{fig-faraday-MAdicaus}
\end{figure}

Although we could try to treat this general case,
 it is instructive to simplify this 
 and take the case of zero current,
 and all $\pXdsurface$ oriented along $z$, 
 and $\pBfieldv$ primarily aligned along,
 and varying in, 
 the $z$ direction,
 so that only its $\pBfield_z$ component is significant.
Subject to these assumptions, 
 and calculating the effective magnetic flux as 
 $\pXmagflux_\textup{A} = \pArea \pBfield_z$, 
 i.e. after moving the RHS surface integral
 through the spatial derivatives,
 we can write a simplified equation
~
\begin{align}
    \pPderivT 
  \pXemf
&=
 -
  \pLightspeed^2
  \pPderivZ^2 \pArea \pBfield_z
=
 -
  \pLightspeed^2
  \pPderivZ^2 \pXmagflux_\textup{A}
.
\label{eqn-MAdicaus}
\end{align}
A causal diagram \cite{Kinsler-2015arXiv-dicaus}
 for the model is shown in fig. \ref{fig-faraday-MAdicaus}.

As a final note, 
 in situations where it is reasonable to 
 replace $c \pPderivZ$ with a $\pPderivT$,
 we can also recover something equivalent to the usual form,
 but with an extra time derivative on both sides.
However, 
 with this replacement the flux $\pXmagflux_\textup{A}$
 is again subjected to the highest order time derivative.
Thus the induction model has also returned to 
 the same causal interpretation --
 one that is reversed from our ``flux as a cause'' preference.

%
\section{Lorentz Force Law}\label{S-forcelaw}

Despite the relatively familiar nature 
 of the  calculations in sections \ref{S-faraday} and \ref{S-ampere},
 in many ways 
 it might be seen as more natural to derive a model of induction of current
 by starting with 
 the Lorentz Force Law for the momentum $\pMomentumv$ of an electric charge $\pCharge$
 in a purely magnetic field:
~
\begin{align}
  \pPderivT
  \pMomentumv
=
  \pForcev
&=
  \pCharge \pVelocity \cross \pBfieldv
.
\label{eqn-LorentzForce}
\end{align}
This is in and of itself
 a model where changes in the charge's momentum
 are caused by the (magnetic) Lorentz force $\pForcev$.

As in the previous calculations, 
 we now hope to find it useful to define an EMF-like quantity.
However, 
 whereas those calculations derived EMF from a line integral
 of the electric field, 
 here we instead base it on an integral
 over charges that are moving;
 and to emphasize this distinction we denote it 
 $\pXemf_L$ rather than simply $\pXemf$.

%
\subsection{A simple loop}
\label{S-forcelaw-simple}

Here we will apply it to (one of) the
 charges in our conducting loop.
This is a more general formulation of that in 
  Bleaney and Bleaney \cite[p.160]{Bleaney2-EM}, 
 but they use it merely as a support for something equivalent
 to the standard calculation as done in section \ref{S-faraday}.

Since $\pXemf_L=\pWork / \pCharge$ for 
 potential (EMF) $\pXemf_L$, 
 work $\pWork$, 
 and charge $\pCharge$, 
 and work increment on a charge
 is given by $d\pWork = \pForcev \cdot \pXdline$ 
 with force $\pForce$ along line element $\pXdline$,
~
\begin{align}
 d\pXemf_L
&=
 \left(
   \pVelocity \cross \pBfieldv
 \right)
 \cdot
 \pXdline
.
\end{align}

First we line-integrate over the loop (contour $\pContour)$
 containing charges of $\pCharge$
 and travelling at $\pVelocity$,
 to get the total EMF $\pXemf_L$,
~
\begin{align}
 \pXemf_L
&=
 \pCharge^{-1}
 \oint_\pContour \pXdline \cdot \pForce 
=
  \oint_\pContour \pXdline \cdot
  \left(
    \pVelocity \cross \pBfieldv
  \right)
.
\end{align}

Next we convert the RHS line integral
 to a surface integral of a curl by Stoke's theorem
~
\begin{align}
 \pXemf_L
&=
 \pCharge^{-1}
 \oint_\pContour \pXdline \cdot \pForce 
=
  \int_\pSurface \pXdsurface \cdot
  \grad
  \cross
  \left(
    \pVelocity \cross \pBfieldv
  \right)
,
\end{align}
 so that now we can use the standard vector identity
 for $\grad \cross (\Vec{A} \cross \Vec{B})$,
~
\begin{align}
 \pXemf_L
&=
  \int_\pSurface \pXdsurface \cdot
  \left[
    \pVelocity \left( \grad \cdot \pBfieldv \right)
   -
    \pBfieldv \left( \grad \cdot \pVelocity \right)
   +
    \left( \pBfieldv \cdot \grad \right) \pVelocity  
   -
    \left( \pVelocity \cdot \grad \right) \pBfieldv
  \right]
.
\end{align}

\UPDATES{The first term vanishes since $\grad \cdot \pBfieldv = 0$;
and we can further simplify 
 by removing the second term 
 if the velocity field for the charges
 (essentially the scaled current) has no sources or sinks.
Here, 
 for convenience, 
 we go further and}
%
 insist on a constant velocity field $\pVelocity$
 so that the third term also vanishes.
The result is
~
\begin{align}
 \pXemf_L
&=
   -
  \int_\pSurface \pXdsurface \cdot
    \left( \pVelocity \cdot \grad \right) \pBfieldv  
.
\label{eqn-Lforce-emf}
\end{align}

\begin{figure}
  \begin{center}
    \resizebox{0.95\columnwidth}{!}{
    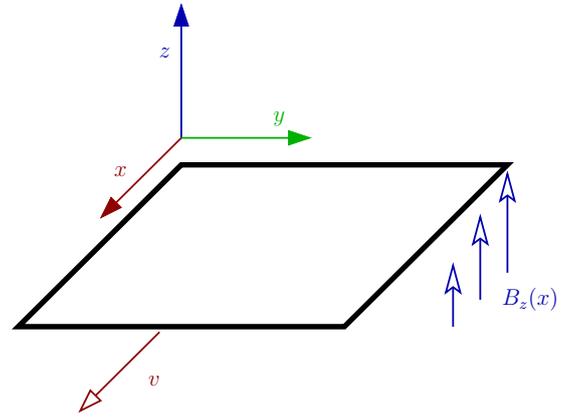}
  \end{center}
\caption{The magnetic field and loop (integration contour)
 orientations used in reducing \eqref{eqn-Lforce-emf}
 to \eqref{eqn-Lforce-emfz}.
The loop in the $x,y$ plane moves with speed $v$
 in the $x$ direction through a $z$-directed magnetic field 
 whose strength varies only with $x$ (i.e. $B_z(x)$).
}
\label{fig-forcecontour}
\end{figure}

To make this result easier to understand, 
 we further simplify the situation
 as shown in fig. \ref{fig-forcecontour}, 
 considering a square loop in the $xy$ plane,
 which 
 bounds the surface $\pSurface$ whose area $\pArea$
 is perpendicular to the $z$ axis.
The magnetic field is oriented along $z$, 
 so that the only non-zero component is $\pBfield_z$, 
 and it varies only along $x$,
 and at a fixed rate.
We then move our loop sideways parallel to the $x$ axis
 at a fixed speed $\pSpeed$, 
 so that 
 we get a straightforward equality, 
 but one without any time derivatives,
 and therefore \emph{no implied causality}:
~
\begin{align}
 \pXemf_L
&=
   -
  \pArea 
  \pSpeed
  \left[ \frac{d \pBfield_z}{dx} \right]
,
\label{eqn-Lforce-emfz}
\end{align}
This seems most closely related to the Maxwell-Ampere formulation
 in section \ref{S-ampere}, 
 which also depends on spatial gradients of $\pBfield$.
However,
 we could choose to adapt this relation
 by converting
 the numerator $dx$ part of the speed $\pSpeed = {dx}/{dt}$,
 and the denominator $dx$ part of the field gradient ${d \pBfield_z}/{dx}$,
 to get 
~
\begin{align}
 \pXemf_L
&=
   -
  \pArea 
  \left[ \frac{dx}{dt} \right]
  \left[ \frac{d \pBfield_z}{dx} \right]
\rightarrow
   -
  \pArea 
  \left[ \frac{d \pBfield_z}{dt} \right]
,
\end{align}
 which is the normal expression of Faraday's Law, 
 {if we assume the equivalence} 
 of $\pXemf_L$ and $\pXemf$,
 \emph{despite their being derived from different starting points}.
Regardless of this, 
 we have still failed to find a mathematical model 
 with the causal properties we are looking for;
 where properties of the magnetic field or its flux 
 cause currents to appear or change.

%
\subsection{{\Ushape}-shaped bar and moving rod}\label{S-forcelaw-example}

\begin{figure}
  \begin{center}
    \resizebox{0.95\columnwidth}{!}{
    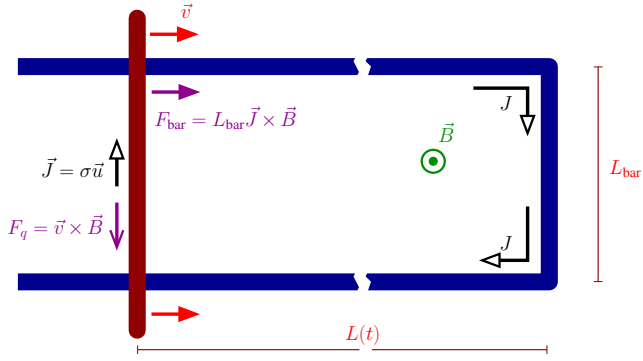}
  \end{center}
\caption{Diagram showing the {\Ushape}-shaped bar in dark blue, 
 and the moving rod in dark red.
Black arrows indicate the current ${\sCurrent}$,  
 and
 red arrows show the (positive) motion of the moving bar.
Magenta arrows indicate the two forces: 
 one on the bar, as caused by the effect of the (its) current travelling 
 through the magnetic field; 
 and the other 
 on the charges in the bar (and hence on the current), 
 caused by the motion of the bar through the magnetic field.
Along the bar,
 it is useful to express the current $\sCurrent$ 
 as a vector ($\vCurrent$), to help with calculating the direction
 of $\Fbarv$.
}
\label{fig-faraday-U}
\end{figure}

Let us now try to 
 make more progress by treating a specific situation
 instead of abstract constructions. 
Consider the standard undergraduate style system of a long rectangular
 conducting {\Ushape}-shape in a constant magnetic field, 
 with a conducting bar closing a current loop, 
 as depicted in fig. \ref{fig-faraday-U}.
The bar then is said to slide along the arms of the {\Ushape},
 thus changing the area enclosed by the loop,
 hence changing the enclosed flux.
We might then straightforwardly apply Faraday's Law 
 from the LHS of \eqref{eqn-inductorizing},
 and use the rate of change of flux to calculate an EMF, 
 and then (if we want to) an induced current associated with that EMF.

However, 
 given our decision to insist on there being 
 a particular causal interpretation,
 and because Faraday's Law does not comply with that interpretation, 
 we must therefore choose a different mathematical model.
Since without assuming spatial gradients of the field, 
 which are not a necessary feature of our example,
 we cannot use the Ampere-based expression in \eqref{eqn-FaradayAmpere},
 we will start with the force law, 
However, 
 using the Lorentz Force Law \eqref{eqn-LorentzForce}
 means that 
 we can see immediately that
 there are two forces on charges relevant here:

\textit{First,}
 there is the
 the magnetic force on the charges in the bar, 
 due to the motion of the bar 
 in the magnetic field.
This force is oriented along the bar, 
 modifying the charge velocity (electric current) in the bar, 
 and then by extension in the entire loop.

\textit{Second,}
 there is a magnetic force on the charges in the bar, 
 due to the along-bar motion of those charges (i.e. the current)
 in the magnetic field.
This force pushes the charges perpendicular to the bar, 
 which can be related to the Hall effect.
Here we assume that this Hall effect rapidly polarizes the bar, 
 on a timescale much faster than other processes,
 so that the net effect shows only in 
 a residual slowing force on the bar itself.

This means that we can use the force law to calculate
 (a) 
 how the current in the loop responds to the velocity ${\vbar}$ of the bar, 
 and 
 (b)
 how the motion of the bar responds to the current $\vCurrent$ in the loop.
These will naturally be dynamic equations with
 an explicit causality.

Here we take the wire {\Ushape} to be fixed and rigid, 
 so that any forces on it due to current flow within it, 
 and/or the presence of electromagnetic fields,
 are neglected.
The wire {\Ushape} itself has arbitrarily long arms
 with a separation and base length of ${\Lbar}$, 
 each with a linear charge density ${\Qdensity}$.
The bar has mass $\Mbar$ and 
 velocity ${\vbar}$,
 to give a momentum ${\Pbar}$.
It is mounted perpendicular to the wire {\Ushape}'s arms, 
 a distance $\Larm$ from their base,  
 and (also) has a charge density ${\Qdensity}$ and length ${\Lbar}$.

The \emph{current} ${\sCurrent}$
 consists of flow of charges with forward momentum ${\Pcur}$.
It has a total charge ${\Qtotal}$, 
 but with only ${\Qbar} = {\Lbar} {\Qdensity}$ in the bar, 
 which is the only part that gets pushed by the magnetic force.
If the (effective) 
 mass density of the charges\footnote{Attempting to describe
 accurately how the charges respond as a group
 is difficult and complicated.
One must somehow average over the many distinct microscopic influences
 on each charge: 
 not only from its local environment in the conductor
 and any externally applied fields, 
 but also the many interactions between it and the other charges
  \cite{Darwin-1936prsa,Essen-2009ejp,Boyer-2015ajp} must be considered.
 Here we simply reduce the combined effect of all such complications
  down into an effective mass density.}
 is ${\QMdensity}$, 
 the velocity of the charges along the bar is $\vcharge$,
 and their speed is $\scharge = {\Pcur} / {\QMdensity} \Ltotal(t)$
 where $\Ltotal(t) = 2 {\Lbar} + 2 \Larm(t)$.
The corners of the {\Ushape} and the
 contacts between the {\Ushape} and the bar
 are assumed to divert (e.g. perhaps by elastic reflection)
 the charges around the loop (i.e. into their new direction), 
 so that ${\Pcur}$ is only either forward 
 (clockwise on fig. \ref{fig-faraday-U})
  or backward (anticlockwise) around the loop; 
 the charge density of the current is also assumed
 to be fixed\footnote{A further detail, 
  neglected here, 
  is that each movement of the bar
  either orphans or inherits
  some extra charges due to the change in side-wide length.}.

The total force on the \emph{bar} is only due to the magnetic force, 
 which needs to be summed over the length ${\Lbar}$.
Note that the field inside the loop is not important here, 
 what counts is its strength at each point in the bar,
 at the instant the total force is integrated.
The force density on each element of the bar is
~
\begin{align}
  \pPderivT \Vec{p}_{bar}
&=
  {\Qdensity} {\vcharge} 
 \cross
  \pBfieldv 
,
\end{align}
~
so that in total we get 
~
\begin{align}
  \pPderivT {\Pbar}
&=
  {\Lbar} {\Qdensity} {\Pcur} \pBfield_z / {\QMdensity} \Ltotal(t)
.
\label{eqn-forcelaw-example-dtPbar}
\end{align}

The total force on the \emph{current} is not only the magnetic force, 
 which needs to be summed over the bar's length ${\Lbar}$, 
 but also includes a linear resistive term $\gamma(t)$, 
 where $\gamma(t) = r \Ltotal(t)$
 and $r$ is proportional to resistance per unit length.
Note that (again) the field inside the loop is not important here, 
 what counts is its strength along the bar
 at its current location.
Since the speed of the bar along the wires is ${\Pbar} / {\Mbar}$, 
 the force density on each element of the current is
~
\begin{align}
  \pPderivT \Vec{p}_{cur}
&=
 -
  {\Qdensity} {\vbar} \cross \pBfieldv
 -
  r \Vec{p}_{cur}
\end{align}
~
 so that in total, 
 we have
~
\begin{align}
  \pPderivT {\Pcur}
&=
 -
  {\Lbar} {\Qdensity} {\Pbar} \pBfield_z / {\Mbar}
 -
  \gamma(t) {\Pcur}
.
\label{eqn-forcelaw-example-dtPcur}
\end{align}

These are two linear coupled differential equations, 
 and can be combined most easily by taking the time derivative
 of the ${\Pcur}$ equation, 
 and substituting one into the other.
Simplifying the equations by not displaying the 
 explicit time dependences of $\Ltotal(t)$ and $\gamma(t)$, 
 we get
~
\begin{align}
  \pPderivT^2 {\Pcur}
&=
  \left( -{\Lbar} {\Qdensity}  \pBfield_z / {\Mbar} \right)
  \pPderivT {\Pbar}
 -
  {\pPderivT} \gamma {\Pcur}
\nonumber
\\
&\qquad
 +
  \left( -{\Lbar} {\Qdensity}  {\Pbar} / {\Mbar} \right)
  \pPderivT \pBfield_z
\\
&=
 -
  \left( {\Lbar} {\Qdensity}  \pBfield_z / {\Mbar} \right)
  \left[
    {\Lbar} {\Qdensity} {\Pcur} \pBfield_z / {\QMdensity} \Ltotal
  \right]
\nonumber
\\
&\qquad
 -
  {\pPderivT} \gamma(t) {\Pcur}
 -
  \left( {\Lbar} {\Qdensity}  {\Pbar} / {\Mbar} \right)
  \pPderivT \pBfield_z
\\
&=
 -
  \frac{ {\Lbar}^2 {\Qdensity}^2  \pBfield_z^2 }
       { {\Mbar} {\QMdensity} \Ltotal }
  {\Pcur}
 -
  {\pPderivT} \gamma {\Pcur}
\nonumber
\\
&\qquad
 -
  \left( {\Lbar} {\Qdensity}  {\Pbar} / {\Mbar} \right)
  \pPderivT \pBfield_z
\\
&=
 -
  \Omega^2
  {\Pcur}
 -
  {\pPderivT} \gamma {\Pcur}
 -
  \left( {\Qdensity}  {\Lbar} {\Pbar} / {\Mbar} \right)
  \pPderivT \pBfield_z
\\
&=
 -
  \Omega^2
  {\Pcur}
 -
  {\pPderivT} \gamma {\Pcur}
 -
  \left( {\Qdensity}  {\Lbar} \sbar \right)
  \pPderivT \pBfield_z
,
\label{eqn-current2ode}
\end{align}
 where 
~
\begin{align}
  \Omega(t)
&=
  \frac{ {\Lbar} {\Qdensity}  \pBfield_z }
       { {\QMdensity} \sqrt{ {\Mbar} \Ltotal(t) } }
.
\end{align}

Note that 
 we cannot directly replace ${\Pcur}$ 
 with current ${\sCurrent}$ in the differential equations, 
 since they are related via the total conductor length $\Ltotal(t)$.
Further, 
 the driving term dependent on $\pPderivT \pBfield_z$
 is proportional to $\sbar(t) = {\Pbar}(t) / {\Mbar}$, 
 which has its own dynamics:
 if the magnetic field is changing in time, 
 it is not sufficient to try to solve only \eqref{eqn-current2ode}, 
 we also need to consider \eqref{eqn-forcelaw-example-dtPbar}.

Since there are some complicated interdependences 
 contained within this result, 
 we now consider some instructive special cases.

\subsubsection{Constant magnetic field}
\label{S-example-Bconst}

Consider the case where the magnetic field is a constant
 over all time as well as space,
 so that $\pPderivT \pBfield_z = 0$; 
 i.e. the third RHS term of \eqref{eqn-current2ode} vanishes.
Here, 
 the bar, 
 after being given an initial push,
 will oscillate forwards and backwards 
 according to the frequency parameter $\Omega(t)$, 
 but with the amplitude of
 those oscillations dying away 
 with a rate given by $\gamma(t)$.

We can also see that for small variations in $\Larm(t)$ and $\Ltotal(t)$,
 the frequency parameter will be effectively constant, 
 and so become a true frequency of oscillation.
In this regime, 
 we can straightforwardly substitute $\Pcur$
 with the current $\sCurrent = \Qdensity \Pcur / \QMdensity \Ltotal$,
 and get
~
\begin{align}
 \pPderivT^2 \pCurrent
&=
 -
  \Omega^2
  \pCurrent
 -
  {\pPderivT} \gamma \pCurrent
.
\end{align}
This would result in damped sinusoidal variations in current, 
 and hence related oscillations in the speed of the bar; 
 since the acceleration of the bar can be related to the current by 
 \eqref{eqn-forcelaw-example-dtPbar}.

\subsubsection{Time varying magnetic field}
\label{S-example-Bt}

The result \eqref{eqn-current2ode} has retained the possibility for a 
 time-dependent $\pBfield_z$.
We can see that the change in the field $\pPderivT \pBfield_z$
 acts to accelerate the current, 
 but in manner dependent on ${\Pbar}$
 (i.e. the bar velocity).
So the current still wants to oscillate, 
 albeit in a way unlikely to be a simple sinusoid, 
 with the field changes acting as a driving term.

Note that causality is maintained,  
 since the effect on ${\Pcur}$ is second order in time, 
 one order greater than that on $\pBfield_z$
 \cite{Kinsler-2011ejp,Kinsler-2010pra-lfiadc}.

\subsubsection{EMF}
\label{S-example-EMF}

We can now consider what \eqref{eqn-current2ode}
 tells us about the EMF induced in the loop; 
 and since whate we are doing is based on the Lorentz force law, 
 we use EMF $\pXemf_L$
 rather than the usual Faraday Law $\pXemf$.
Since $\pXemf_L$ is the work per unit charge, 
 or the line (loop) integral of the force, 
 and force is just the change in the momentum, 
 we have that $\pXemf_L = \Qdensity^{-1} \pPderivT \Pcur$.
This means that for small changes in $\Ltotal(t)$, 
 we can almost reuse \eqref{eqn-current2ode} directly --
 we just apply another time derivative and substitute for $\Pcur$, 
 to get 
~
\begin{align}
  {\pPderivT}^2 \pXemf_L
&=
 -
  \Omega^2
  \pXemf_L
 -
  {\pPderivT} \gamma
  \pXemf_L
\nonumber
\\
&\qquad
 -
  \pPderivT
  \left[
    \left( {\Lbar}  \sbar \right)
    \pPderivT \pBfield_z
  \right]
.
\label{eqn-emf2ode}
\end{align}

The equation might at a first glance
 appear to show time-like changes in the magnetic field
 causing alterations in the EMF.
Of course, 
 since it contains more model dependent detail, 
 some differences with 
 standard Faraday's Law in \eqref{eqn-induction}
 or the alternate Ampere-derived one in \eqref{eqn-FaradayAmpere}
 are to be expected.

The difficulty now is that the right hand side, 
 as a direct result of the conversion from ${\Pcur}$ to $\pXemf_L$
 now has a second order time derivative acting on $\pBfield$, 
 just as does the left hand side acting on $\pXemf_L$.
Since they have equal orders of time derivative, 
 they share the same causal status --
 combined together correctly they could be interpreted 
 as being the effect of the other terms in \eqref{eqn-emf2ode}, 
 i.e. the cause-like terms $\pXemf_L$ and ${\pPderivT} \gamma \pXemf_L$, 
 as well as ones involving $\sbar$.

This means that \eqref{eqn-emf2ode}
does not have the causal interpretation we were searching for, 
 e.g. such as 
 where first-order time-like changes in the magnetic field
 caused second-order alterations in the EMF.

\subsection{Discussion}
\label{S-forcelaw-discussion}

Here we have seen that for both our force law based calculations, 
 i.e.
 the very simple one 
 and  
 the more realistic system, 
 both fail to supply us with a result where 
 properties of the magnetic field can be said to cause changes in the EMF.
Nevertheless, 
 the causal attributions like that which we aimed for can still be made by
 not referring to the EMF, 
 but instead using 
 statements like 
 (changes in) magnetic field properties induce \emph{currents}.

%
\section{Conclusion}\label{S-conclusion}

In this paper we have tried to search for
 a version or rederivation of Faraday's Law 
 whose mathematical form mimics the causal interpretation
 we would like to have, 
 namely:
 changes in magnetic flux through a loop 
 induce changes in the EMF around that loop.
However, 
 the usual mathematical form of Faraday's Law 
 is incompatible with this desire, 
 and only allows us to say that EMF causes changes in flux.

To address this apparent deficiency,
 we derived a Faraday-like law based on the Maxwell Ampere curl equation, 
 which indeed allowed us to talk of induced changes in the EMF, 
 but these were instead caused by
 spatial gradients in the magnetic flux, 
 which is not quite what we had hoped.
This led us to try 
 an alternative approach based on the Lorentz force law, 
 whose microscopic foundations
 showed promise in that properties of the magnetic field
 indeed induced (caused) changes in current, 
 both in an abstract 
 and a more realistic setting.
However, 
 when the mathematical model was converted to 
 refer instead to induced changes in an EMF-like quantity $\pXemf_L$,
 we found that the model became incompatible with
 the desired causal interpretation.

In summary,  
 our investigation has shown that one should be careful
 when making causal interpretations of magnetic induction processes.
First, 
 one should clearly distinguish between interpretations relevant to 
 an experiment, 
 and those relevant to a mathematical model; 
 a distinction that is vital in the case of Faraday's Law.
Second, 
 since the closest we can get to a 
 ``changes in magnetic flux induces EMF'' is based on
 the Maxwell Ampere equation and not Maxwell Faraday, 
 we need to be careful when deciding on causal interpretations
 of empirical laws derived from or compatible with experiment.

\section*{Acknowledgements}

I am grateful for the support provided
 by STFC (the Cockcroft Institute ST/G008248/1
 and ST/P002056/1)
 and EPSRC (the Alpha-X project EP/N028694/1). 
I would also like to acknowledge the hospitality of Imperial College London, 
 and many interesting discussions with 
 Jonathan Gratus and
 Martin W. McCall;
 as well as other debate on \url{news:sci.physics.electromag}.

%

\begin{thebibliography}{15}
\expandafter\ifx\csname natexlab\endcsname\relax\def\natexlab#1{#1}\fi
\expandafter\ifx\csname bibnamefont\endcsname\relax
  \def\bibnamefont#1{#1}\fi
\expandafter\ifx\csname bibfnamefont\endcsname\relax
  \def\bibfnamefont#1{#1}\fi
\expandafter\ifx\csname citenamefont\endcsname\relax
  \def\citenamefont#1{#1}\fi
\expandafter\ifx\csname url\endcsname\relax
  \def\url#1{\texttt{#1}}\fi
\expandafter\ifx\csname urlprefix\endcsname\relax\def\urlprefix{URL }\fi
\providecommand{\bibinfo}[2]{#2}
\providecommand{\eprint}[2][]{\url{#2}}

\bibitem[{\citenamefont{Bleaney and Bleaney}(1965)}]{Bleaney2-EM}
\bibinfo{author}{\bibfnamefont{B.~I.} \bibnamefont{Bleaney}} \bibnamefont{and}
  \bibinfo{author}{\bibfnamefont{B.}~\bibnamefont{Bleaney}},\\
  \emph{\bibinfo{title}{Electricity and Magnetism}}\\
  (\bibinfo{publisher}{Clarendon Press}, \bibinfo{address}{Oxford},
  \bibinfo{year}{1965}), \bibinfo{edition}{2nd} ed.

\bibitem[{\citenamefont{Bohren}(2010)}]{Bohren-2010ejp}
\bibinfo{author}{\bibfnamefont{C.~F.} \bibnamefont{Bohren}},
  \\ \bibinfo{journal}{Eur. J. Phys.} \textbf{\bibinfo{volume}{31}},
  \bibinfo{pages}{573} (\bibinfo{year}{2010}),
  \\ \XDOI{10.1088/0143-0807/31/3/014}.

\bibitem[{\citenamefont{Landau and Lifshitz}(1984)}]{LandauLifshitz}
\bibinfo{author}{\bibfnamefont{L.~D.} \bibnamefont{Landau}} \bibnamefont{and}
  \bibinfo{author}{\bibfnamefont{E.~M.} \bibnamefont{Lifshitz}},\\
  \emph{\bibinfo{title}{Electrodynamics of Continuous Media}}\\
  (\bibinfo{publisher}{Pergamon}, \bibinfo{address}{Oxford and New York},
  \bibinfo{year}{1984}).

\bibitem[{\citenamefont{Lucarini et~al.}(2005)\citenamefont{Lucarini, Saarinen,
  Peiponen, and Vartiainen}}]{KKinOMR}
\bibinfo{author}{\bibfnamefont{V.}~\bibnamefont{Lucarini}},
  \bibinfo{author}{\bibfnamefont{J.~J.} \bibnamefont{Saarinen}},
  \bibinfo{author}{\bibfnamefont{K.~E.} \bibnamefont{Peiponen}},
  \bibinfo{author}{\bibfnamefont{E.~M.}
  \bibnamefont{Vartiainen}}, \\
 \emph{\bibinfo{title}{Kramers-Kronig Relations in Optical Materials Research}}, \\
  vol. \bibinfo{volume}{110}, 
  (\bibinfo{publisher}{Springer}, \bibinfo{address}{Berlin / Heidelberg},
  \bibinfo{year}{2005}), 
  \\ ISBN \bibinfo{isbn}{978-3-540-23673-3},
  \\ \XDOI{10.1007/b138913}.

\bibitem[{\citenamefont{Kinsler}(2011)}]{Kinsler-2011ejp}
\bibinfo{author}{\bibfnamefont{P.}~\bibnamefont{Kinsler}},
  \\ \bibinfo{journal}{Eur. J. Phys.} \textbf{\bibinfo{volume}{32}},
  \bibinfo{pages}{1687} (\bibinfo{year}{2011}), \\
  \XDOI{10.1088/0143-0807/32/6/022};\\
 \bibinfo{note}{see also the updated version at \XARXIV{1106.1792} 
  \cite{Kinsler-2011arxiv-how2c}}.

\bibitem[{\citenamefont{Corduneanu}(2002)}]{Corduneanu-FECAUSAL}
\bibinfo{author}{\bibfnamefont{C.} \bibnamefont{Corduneanu}},\\
  \emph{\bibinfo{title}{Functional Equations with Causal Operators}}\\
  (\bibinfo{publisher}{Taylor and Francis}, \bibinfo{address}{London},
  \bibinfo{year}{2002}).

\bibitem[{\citenamefont{Corduneanu}(2000)}]{Corduneanu-2000jde}
\bibinfo{author}{\bibfnamefont{C.}~\bibnamefont{Corduneanu}},
  \\ \bibinfo{journal}{J. Differ. Equations} \textbf{\bibinfo{volume}{168}},
  \bibinfo{pages}{93--101} (\bibinfo{year}{2000}), \\
  \XDOI{10.1006/jdeq.2000.3879}.

\bibitem[{\citenamefont{Kinsler}(2011)}]{Kinsler-2011arxiv-how2c}
  \bibinfo{author}{\bibfnamefont{P.}~\bibnamefont{Kinsler}},
  \\ \bibinfo{note}{``How to be causal''}, 
  \\ \XARXIV{1106.1792}

\bibitem[{\citenamefont{Kinsler}(2014)}]{Kinsler-2014arXiv-negfreq}
\bibinfo{author}{\bibfnamefont{P.}~\bibnamefont{Kinsler}}
  (\bibinfo{year}{2014}), \\
  \bibinfo{note}{``What I
  talk about when I talk about propagation''}, \\
  \XARXIV{1408.0128}.

\bibitem[{\citenamefont{Kinsler}(2018)}]{Kinsler-2018jo-d2owe}
\bibinfo{author}{\bibfnamefont{P.}~\bibnamefont{Kinsler}},
  \\ \bibinfo{journal}{J. Opt.} \textbf{\bibinfo{volume}{20}},
  \bibinfo{pages}{025502} (\bibinfo{year}{2018}), \\
  \XDOI{10.1088/2040-8986/aaa0fc}.

\bibitem[{\citenamefont{Rideout and Sorkin}(2000)}]{Rideout-S-2000prd}
\bibinfo{author}{\bibfnamefont{D.~P.}~\bibnamefont{Rideout}},
\bibinfo{author}{\bibfnamefont{R.~D.}~\bibnamefont{Sorkin}},
  \\ \bibinfo{journal}{Phys. Rev. D} \textbf{\bibinfo{volume}{61}},
  \bibinfo{pages}{024002} (\bibinfo{year}{2000}), \\ \XARXIV{gr-qc/9904062},
  \\ \XDOI{10.1103/PhysRevD.61.024002}.

\bibitem[{\citenamefont{Jefimenko}(2004)}]{Jefimenko-2004ejp}
\bibinfo{author}{\bibfnamefont{O.}~\bibnamefont{Jefimenko}},
  \\ \bibinfo{journal}{Eur. J. Phys.} \textbf{\bibinfo{volume}{25}},
  \bibinfo{pages}{287} (\bibinfo{year}{2004}),
  \\ \XWEB{http://iopscience.iop.org/0143-0807/25/2/015}.

\bibitem[{\citenamefont{Hill}(2010)}]{Hill-2010pt}
\bibinfo{author}{\bibfnamefont{S.~E.} \bibnamefont{Hill}},
  \\ \bibinfo{journal}{The Physics Teacher} \textbf{\bibinfo{volume}{48}},
  \bibinfo{pages}{410} (\bibinfo{year}{2010}),
  \\ \XDOI{10.1119/1.3479724}.

\bibitem[{\citenamefont{Savage}(2012)}]{Savage-2011arXiv-cem}
\bibinfo{author}{\bibfnamefont{C.~M.} \bibnamefont{Savage}},
  \\ \bibinfo{journal}{The Physics Teacher} \textbf{\bibinfo{volume}{50}},
  \bibinfo{pages}{226} (\bibinfo{year}{2012}),
  \\ \XDOI{10.1119/1.3694075},
  \\ \XARXIV{1105.1197}.

\bibitem[{\citenamefont{Boyer}(2015)}]{Boyer-2015ajp}
\bibinfo{author}{\bibfnamefont{T.~H.} \bibnamefont{Boyer}},
  \\ \bibinfo{journal}{Am. J. Phys.} \textbf{\bibinfo{volume}{83}},
  \bibinfo{pages}{263} (\bibinfo{year}{2015}),
  \\ \XDOI{10.1119/1.4901191}.

\bibitem[{\citenamefont{Thide}(2004)}]{Thide-EMFT}
\bibinfo{author}{\bibfnamefont{B.}~\bibnamefont{Thide}},\\
  \emph{\bibinfo{title}{Electromagnetic Field Theory}}\\
  (\bibinfo{publisher}{Upsilon Books}, \bibinfo{address}{Uppsala, Sweden},
  \bibinfo{year}{2004}).

\bibitem[{\citenamefont{Jackson}(1999)}]{Jackson-ClassicalED}
\bibinfo{author}{\bibfnamefont{J.~D.} \bibnamefont{Jackson}},
  \emph{\bibinfo{title}{Classical Electrodynamics}}
  (\bibinfo{publisher}{Wiley}, \bibinfo{year}{1999}), \bibinfo{edition}{3rd}
  ed., ISBN \bibinfo{isbn}{978-0-471-30932-1}.

\bibitem[{\citenamefont{Reitz et~al.}(1980)\citenamefont{Reitz, Milford, and
  Christy}}]{RMC}
\bibinfo{author}{\bibfnamefont{J.~R.} \bibnamefont{Reitz}},
  \bibinfo{author}{\bibfnamefont{F.~J.} \bibnamefont{Milford}},
  \bibnamefont{and} \bibinfo{author}{\bibfnamefont{R.~W.}
  \bibnamefont{Christy}}, \\
  \emph{\bibinfo{title}{Foundations of electromagnetic theory}} \\
  (\bibinfo{publisher}{Addison-Wesley}, \bibinfo{year}{1980}),
  \bibinfo{edition}{3rd} ed.

\bibitem[{\citenamefont{Kinsler}(2015)}]{Kinsler-2015arXiv-dicaus}
\bibinfo{author}{\bibfnamefont{P.}~\bibnamefont{Kinsler}},
 (\bibinfo{year}{2015}), \\
  \bibinfo{note}{``Causal diagrams for physical models''}, \\
  \XARXIV{1509.01491}. 

\bibitem[{\citenamefont{Kumar}(2017)}]{Kumar-2017ejp}
\bibinfo{author}{\bibfnamefont{V.}~\bibnamefont{Kumar}}, \\ 
 \bibinfo{journal}{Eur. J. Phys.} 
 \textbf{\bibinfo{volume}{38}}, 045203 (\bibinfo{year}{2017}),
  \\ \XDOI{10.1088/1361-6404/aa6cea}.

\bibitem[{\citenamefont{Darwin}(1936)}]{Darwin-1936prsa}
\bibinfo{author}{\bibfnamefont{C.~G.}~\bibnamefont{Darwin}},
  \\ \bibinfo{journal}{Proc. Roy. Soc. A} \textbf{\bibinfo{volume}{154}},
  \bibinfo{pages}{61--66} (\bibinfo{year}{1936}), 
  \\ \XDOI{10.1098/rspa.1936.0036}.

\bibitem[{\citenamefont{Essen}(2009)}]{Essen-2009ejp}
\bibinfo{author}{\bibfnamefont{H.}~\bibnamefont{Essen}}, \\ 
 \bibinfo{journal}{Eur. J. Phys.} 
 \textbf{\bibinfo{volume}{30}} (\bibinfo{year}{2009}),
  \\ \XDOI{10.1088/0143-0807/30/3/009}.

\bibitem[{\citenamefont{Kinsler}(2010)}]{Kinsler-2010pra-lfiadc}
\bibinfo{author}{\bibfnamefont{P.}~\bibnamefont{Kinsler}},
  \\ \bibinfo{journal}{Phys. Rev. A} \textbf{\bibinfo{volume}{82}},
  \bibinfo{pages}{055804} (\bibinfo{year}{2010}), \\ \XARXIV{1008.2088},
  \\ \XDOI{10.1103/PhysRevA.81.013819}.




\end{thebibliography}

%

\section*{Appendix: Commentary}

Many treatments of induction involve both 
 discussion of the experimental observations 
 as well as 
 a mathematical model (Faraday's Law)
 used to describe the process.
This combination of topics, 
 whilst being a perfectly natural one, 
 often makes it hard to pin down what attributions of causality
 are being made.
Notably,
 when reading about magnetic induction, 
 and
 on finding a statement along the lines of 
 ``a change in magnetic flux causes (or induces) an EMF'',
 it is frequently not clear whether it refers to:

\begin{description}

\item[(a)]
 an inference made on the 
 basis of experimental observation, 
 where it may be perfectly reasonable.

\item[(b)]
 an inference based on Faraday's Law
 (i.e. \eqref{eqn-inductorizing}), 
 where it is emphatically not justified

\end{description}

Ironically, 
 therefore, 
 it is often the more mathematical treatments 
 which are an easy target for criticism, 
 since their statements tend to refer almost unambiguously
 to the mathematical model (e.g. \cite{Thide-EMFT}).
However, 
 a preceeding statement referring to experiment would most likely
 have been enough to obscure the specific detail
 of the claimed cause-and-effect reasoning.
Indeed, 
 more generally the context of the ``caused'' or ``induced'' claim 
 is even less clear --
 e.g. I own a text \cite{RMC} that makes both the weaker ``associated'', 
 as well as the stronger ``induced EMF'' claim ...
 but against a backdrop where 
 Faraday's Law is carefully introduced as an \emph{experimental} law, 
 but is nevertheless in the presence of mathematical description.

Along these lines, 
 a great deal of simpler material 
 (e.g. at cliffsnotes.com,
  or schoolphysics.co.uk, 
  or hyperphysics.phy-astr.gsu.edu)
 also merges discussions of experiment, 
 theory, 
 and practical calculation.
Therefore 
 one cannot say unequivocally that they are ``wrong'', 
 even if a reader, 
 led by the experimental attribution of causality, 
 has been led to assume that Faraday's Law, 
 i.e. \eqref{eqn-inductorizing},
 is a good model for the 
 cause-and-effect present in the experiment.

For example, 
 imagine we read a treatment that first says
 that in an experiment ``changes in flux induce (cause) EMF'', 
 but then subsequently states or implies
 that this is somehow encapsulated or represented
 by Faraday's Law \eqref{eqn-inductorizing}.
In that case, 
 why should we not --
 naively but wrongly -- 
 then also be led to attribute the same causality
 that applied to an experiment,
 to the mathematical Faraday's Law,
 where it does not apply?
We could very probably defend any such a discussion as not being wrong, 
 but could we really say that it is sufficently clear?

\end{document}